\documentclass{article}
\usepackage[dvips]{graphics}
\usepackage{latexsym}
\begin{document}
\date{}
\title{Dynamics of magnetic Bianchi VI$_0$ cosmologies}
\author{Marsha Weaver$\,$\footnote{Department of Physics,
University of Oregon, Eugene, Oregon 97403}
$\:$\footnote{Present address:
Max-Planck-Institut f\"ur Gravitationsphysik,
Am M\"uhlenberg 1, D-14424 Golm, Germany}
$\:$\footnote{E-mail address:
weaver@aei-potsdam.mpg.de}}
\maketitle
\begin{abstract}
Methods of dynamical systems analysis are used to show rigorously
that the presence of a magnetic field orthogonal to the two
commuting Killing vector fields in any spatially homogeneous
Bianchi type~VI$_0$ vacuum solution to Einstein's equation changes
the evolution toward the singularity from convergent to
oscillatory.  In particular, it is shown that the $\alpha$-limit
set (for time direction that puts the singularity in the past) of
any of these magnetic solutions contains at least two sequential
Kasner points of the BKL sequence and the orbit of the transition
solution between them.  One of the Kasner points in the
$\alpha$-limit set is non-flat, which leads to the result that
each of these magnetic solutions has a curvature singularity.
\end{abstract}


\section{Introduction}

Progress in the characterization of the asymptotic behavior
of spatially homogeneous solutions to Einstein's equation has
been made by the use of methods from dynamical systems analysis.
(See~\cite{book} for an overview.)  For example, these methods
have been used to show that solutions in various spatially
homogeneous subclasses are convergent or oscillatory in a given
time direction.  The vacuum Bianchi type~VI$_0$ spacetimes are 
a class of spatially homogeneous solutions which are known to
be convergent in both time directions~\cite{GD}.  One time
direction is singular and the other is not.  In these spacetimes
the three linearly independent Killing vector fields can be
chosen so that two of them commute with each other.  In
\cite{LKW} evidence is presented that provides strong support
for the conjecture that if there is a magnetic field present
which is orthogonal to the two commuting Killing vector
fields then the approach to the singularity is no longer
convergent but instead oscillatory, and moreover, generically
mixmaster.  This is corroborated by the numerical study
reported in \cite{BGS}.

Rigorous results concerning mixmaster dynamics have been elusive.
Classes of spatially homogeneous solutions believed to be
mixmaster are Bianchi type~IX, Bianchi type~VIII and magnetic
Bianchi type~VI$_0$, with or without a perfect
fluid in each case.  (See \cite{lb1, lb2} for additional
mixmaster classes.)  Not only has the occurrence of mixmaster
dynamics remained a conjecture, but it has not even been shown
that solutions in the classes thought to be mixmaster are
oscillatory.  Progress toward such a result for vacuum Bianchi
type~IX and vacuum Bianchi type~VIII was made in \cite{GD}.  The
magnetic Bianchi type~VI$_0$ spacetimes considered in \cite{LKW}
are considered again in the present paper (without a perfect
fluid) and, building on the results of \cite{LKW} by using
techniques developed in \cite{GD}, it is rigorously shown
that all of them are oscillatory in the singular time direction.

The methods used to show this are those of dynamical systems
analysis.  A discussion of these methods can be found
in \cite{book} and in references cited therein.  In particular,
the formulation of the magnetic Bianchi~VI$_0$ spacetimes
as a dynamical system is taken from \cite{LKW}, and definitions
and results from \cite{LKW} and \cite{GD} are used throughout.
Three types (convergent, oscillatory, mixmaster) of asymptotic
behavior are discussed in section~2.  A review of results from
\cite{LKW} is given in section~3.  The proof of the new result
that magnetic Bianchi~VI$_0$ solutions are oscillatory is
presented in section~4.  A method which might lead to a proof
of mixmaster behavior in this class of spacetimes is suggested
in section~5.

Knowledge of the asymptotic behavior of a solution can lead to
an answer to the question of whether or not an extension through
the initial singularity is possible.  In section~6 it is shown
that each of these solutions has a curvature singularity.
Therefore no extension is possible.  In the course of the
analysis, several other types of spatially homogeneous
solutions to Einstein's equation are considered.  Which of
these have curvature singularities is also determined in
section~6.  The solutions considered here which can be extended
past the initial singularity, and therefore do not have a
curvature singularity, include rotationally symmetric solutions
of Bianchi type~I with a magnetic field orthogonal to the
plane of rotational symmetry.  An extension of these spacetimes
is given in the appendix.  Section~7 consists of concluding remarks.

\section{Types of asymptotic behavior}

In a spatially homogeneous solution to Einstein's equation
there is a foliation by homogeneous spatial hypersurfaces
with time coordinate $t$ constant on each hypersurface.
Assume that the foliation of homogeneous spatial hypersurfaces
is maximally extended.  Choose $t$ so that $t$ takes on all
real values.  Define a generalized Kasner exponent, $p_i(t)$,
to be an eigenvalue of the extrinsic curvature of the
homogeneous hypersurface at time $t$ divided by the mean
curvature of the homogeneous hypersurface~\cite{R1}.  In a
Kasner (vacuum Bianchi type~I)
spacetime the generalized Kasner exponents are exactly
the Kasner exponents and they are constant in time.  There are
other spatially homogeneous solutions to Einstein's equation,
for example the spatially flat Friedmann-Robertson-Walker models,
whose generalized Kasner exponents are constant in time, but
do not satisfy the Kasner relation, $p_1^2 + p_2^2 + p_3^2 = 1$.
A convergent solution (in a particular time direction) is
one in which each of the three generalized Kasner exponents
converges to a limit in that time direction~\cite{R1}.  For
example, if a solution is convergent in the negative time
direction, then $\lim_{t \rightarrow -\infty} p_i(t)$ exists.
A solution may be convergent in both time directions, in
only one time direction, or in neither.

When a class of spatially homogeneous solutions is studied
as a dynamical system using the expansion normalized variables
of Wainwright and collaborators~\cite{book} then a convergent
solution is one whose $\alpha$-limit set (for the negative
time direction, or $\omega$-limit set for the positive time
direction) consists of a single equilibrium point of the
system.  Such an equilibrium point may be a Kasner point,
which represents a Kasner solution,
or it may represent another solution whose generalized Kasner
exponents are constant so that, as in the Kasner solutions,
the evolution is completely characterized by the volume
expansion.  An effect of the normalization by the volume
expansion is that each equilibrium point corresponds to
an entire spacetime and each one-dimensional orbit
corresponds to a one-parameter family of conformally related
spacetimes~\cite{WH}.  Since the volume expansion is zero
at the moment of maximum volume in spacetimes of Bianchi
type~IX, the expansion normalized variables represent half
of the evolution -- either the expanding phase or the collapsing
phase, and so the limit set of a solution in the direction
of maximum expansion gives information about the evolution at
the moment of maximum expansion.  A magnetic Bianchi
type~VI$_0$ spacetime is either always expanding or always
collapsing, depending on choice of time direction, so the
expansion normalized variables cover the entire spatially
homogeneous evolution.

An oscillatory solution is one whose $\alpha$-limit set (or
$\omega$-limit set) contains at least two equilibrium points
of the system.  In all known cases, the equilibrium
points contained in such a limit set are a finite number of
isolated Kasner points and the limit set is a heteroclinic
cycle.  A heteroclinic cycle is a set of $n$ distinct
equilibrium points, $q_i$, and $n$ orbits of the dynamical
system, $\Gamma_i$, such that each orbit is convergent
in both time directions and the $\omega$-limit set of
$\Gamma_i$ consists of the point $q_{i}$ while if $i < n$
the $\alpha$-limit set of $\Gamma_i$ consists of the
point $q_{i+1}$ and if $i = n$ the $\alpha$-limit set
of $\Gamma_i$ consists of the point $q_1$.

Qualitative analysis and the results of numerical simulations
give the following picture for the evolution toward the
singularity in generic solutions belonging to the classes
of spacetimes believed to be mixmaster.  Let $S$ be a
nonexceptional solution in one of these classes.  A sequence
of Kasner evolutions, $q_i$,  approximates the evolution of
$S$.  Given $q_i$, then $q_{i+1}$ is determined by the BKL
map~\cite{BKL1,BKL2,KL,M}. Moreover, when the evolution of
$S$ is not approximately Kasner, during the transition from
$q_i$ to $q_{i+1}$, the evolution of $S$ is approximated by a
spatially homogeneous solution to Einstein's equation which is
uniquely determined by $q_i$. In the case of Bianchi types~VIII
and IX this is a vacuum Bianchi type~II solution.  In the case
of magnetic Bianchi type~VI$_0$ some of the transitions are
approximated by vacuum solutions of Bianchi type~II and some are
approximated by magnetic solutions of Bianchi type~I~\cite{LKW}.
The solution approximating the evolution during the transition
from one Kasner to the next converges to $q_i$ to the future
and to $q_{i+1}$ to the past (for time direction that puts the
singularity in the past).  As the singularity is approached
the approximation of the evolution of $S$ by such a sequence of
Kasners along with the transition solutions determined by the
sequence improves.  However, it is not necessarily expected that
the evolution of $S$ converges to a single BKL sequence of Kasners
along with the transition solutions determined by the sequence
\cite{GD}.  It may be that the evolution always eventually diverges
from one sequence and another one becomes a better approximation.

The description of mixmaster evolution just given can be stated
in terms of the $\alpha$-limit set of $S$.  The presence of a
perfect fluid does not significantly alter the picture.  If
there is no perfect fluid, then each mixmaster class is a
four-dimensional invariant set of a dynamical system and the
solution, $S$, is represented by an orbit in the four-dimensional
invariant set.  The boundary of this four-dimensional set is in
each case the union of the three-dimensional sets listed below
and the compact two-dimensional invariant set, $\cal A$,
consisting of the Kasner points along with the orbits of all
transition solutions between sequential Kasner points of the BKL
sequence.  The boundary of vacuum Bianchi type~IX consists of
orbits of vacuum Bianchi types~VII$_0$ (a three-dimensional set),
II (a two-dimensional set) and I (the Kasners).  In this case
$\cal A$ is the union of the latter two sets.  The boundary
of vacuum Bianchi type~VIII consists of orbits of vacuum Bianchi
types~VII$_0$, VI$_0$ (a three-dimensional set), II and I.
$\cal A$ is again the union of the latter two sets.  The
boundary of magnetic Bianchi type~VI$_0$ consists of orbits
of vacuum Bianchi type~VI$_0$, magnetic solutions of Bianchi
types~II (a three-dimensional set) and I (a two-dimensional
set) and vacuum Bianchi types~II and I.  In this case
$\cal A$ is the union of the latter three sets.  It is known
that the $\alpha$-limit set of $S$ must be contained in the
boundary of the four-dimensional invariant set in which it
lies~\cite{LKW,WH}.  If $S$ is magnetic Bianchi type~VI$_0$
then it is known that its $\alpha$-limit set is non-empty
since the orbit is contained in a compact set of the dynamical
system.  If $S$ is Bianchi type~VIII or IX then until recently
it has not been known that its $\alpha$-limit set is non-empty.
(It appears that this is no longer the case, due
to a recent result by Ringstr\"{o}m showing that such an
$\alpha$-limit set is indeed non-empty \cite{ring}.)
To be consistent with the picture
of mixmaster dynamics, the $\alpha$-limit set of $S$ should
be non-empty, and it should be contained in just part of the
boundary, obtained by excluding the three-dimensional sets.
That is, it should be contained in the set $\cal A$.  This is
because in order for the approximation of the evolution of $S$
by a BKL sequence of Kasners along with the transition solutions
determined by the Kasners in the sequence to get better as the
singularity is approached, the orbit of $S$ must eventually
remain within any compact set containing a neighborhood of
$\cal A$.  Given any point, $p$, of the boundary that is not
in $\cal A$, such a compact set can be found which does not
contain $p$.  Furthermore, if it could be shown that the
$\alpha$-limit set of $S$ is contained in the set $\cal A$,
then the results in \cite{GD} and in the present paper would
show that an entire BKL sequence of Kasner points along
with the transition solutions between them must lie in the
$\alpha$-limit set of $S$.  Whether $\cal A$ or a subset
of $\cal A$ is the $\alpha$-limit set of generic solutions
in the mixmaster classes would still be an open question.

\section{Magnetic Bianchi VI$_0$}

The union of the orbits of magnetic Bianchi~VI$_0$ solutions is the
four-dimensional open invariant set, $U$, defined by~\cite{LKW}:
\begin{equation}
\label{hamcon}
\Sigma_+^2 + \Sigma_-^2 + N_-^2 + {3 \over 2} H^2 = 1,
\end{equation}
\begin{equation}
\label{inequalities}
N_- > 0, \; \; \; N_+^2 < 3 N_-^2, 
\; \; \; \mbox{and} \; \; \; H > 0.
\end{equation}
This formulation is in terms of the shear tensor, the structure
constants and the magnetic field in an orthonormal frame which is
invariant under the action of the group of isometries.  $H$ is
obtained from the magnetic field, which is orthogonal to the two
commuting killing fields.  $\Sigma_+$ and $\Sigma_-$ are obtained
from the shear tensor while $N_+$ and $N_-$ are obtained from the
structure constants.  These quantities have been normalized by
the expansion scalar.  Equation~\ref{hamcon} is the Hamiltonian
constraint.  It is equation (2.19) in \cite{LKW} with $\Omega = 0$.
See \cite{LKW} for further details concerning this construction
and its representation of all solutions in this class.

The time direction is chosen so that the singularity is in
the past ($\tau$ decreasing).  The evolution equations are:
\begin{eqnarray}
\label{equations}
\Sigma_+' & = & - 2 N_-^2 ( 1 + \Sigma_+) + {3 \over 2} H^2 ( 2 -
\Sigma_+), \nonumber \\
\Sigma_-' & = & - ( 2 N_-^2 + {3 \over 2} H^2)
\Sigma_- - 2 N_+ N_-, \nonumber \\
N_+' & = & (2 \Sigma_+ (1 + \Sigma_+) + 2 \Sigma_-^2 + {3 \over 2}
H^2) N_+ + 6 \Sigma_- N_-, \\
N_-' & = & (2 \Sigma_+ (1 + \Sigma_+) + 2 \Sigma_-^2 + {3 \over 2}
H^2) N_- + 2 \Sigma_- N_+, \nonumber \\
H' & = & -(\Sigma_+ (2 - \Sigma_+) - \Sigma_-^2 + N_-^2) H. \nonumber
\end{eqnarray}
These equations can be obtained from equations (2.17) in \cite{LKW}
by using their equation (2.16) with $\Omega = 0$.  Evolution satisfying
equations~\ref{equations} preserves the Hamiltonian constraint.

There is exactly one equilibrium point in $U$, the point
at which $\Sigma_+ = - {1 \over 4}$, $\Sigma_- = 0$,
$N_+ = 0$, $N_- = {3 \over 4}$ and $H = {1 \over 2}$.
In \cite{LKW} it is shown that the $\omega$-limit set of
each solution in $U$ consists of this single equilibrium
point.  That is, all of these magnetic Bianchi~VI$_0$
solutions are convergent in the non-singular time direction.

\begin{figure}
\includegraphics{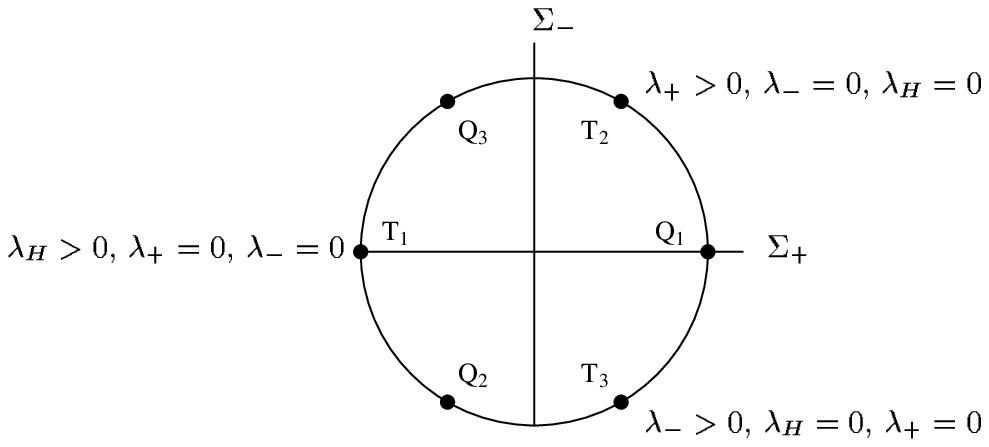}
\caption{The Kasner circle.  The arc between
$T_2$ and $T_3$ containing $Q_1$ is ${\cal K}_1$.  On ${\cal K}_1$,
$\lambda_H < 0$, while $\lambda_+ > 0$ and $\lambda_- > 0$.
The arc between
$T_3$ and $T_1$ containing $Q_2$ is ${\cal K}_2$.  On ${\cal K}_2$,
$\lambda_+ < 0$, while $\lambda_- > 0$ and $\lambda_H > 0$.
The arc between
$T_1$ and $T_2$ containing $Q_3$ is ${\cal K}_3$.  On ${\cal K}_3$,
$\lambda_- < 0$, while $\lambda_H > 0$ and $\lambda_+ > 0$.}
\label{kasnercircle}
\end{figure}

The boundary of $U$, $\partial U$, is made up of various sets
which are invariant with respect to equations~\ref{equations}.
Solutions to equations~\ref{equations} in these invariant sets
represent solutions to Einstein's equation in various other
spatially homogeneous classes.  As mentioned previously, these
are vacuum Bianchi~VI$_0$ ($H=0$), magnetic Bianchi~II
($N_+^2 = 3 N_-^2$) and I ($N_- = 0$), vacuum Bianchi~II
($H=0$ and $N_+^2 = 3 N_-^2$) and the Kasner points ($H=0$ and
$N_- = 0$).  Note that the magnetic field is not the most
general one possible.  See \cite{lb1, lb2} for other possibilities.
Each Kasner point is an equilibrium point and there are no other
equilibrium points in $\partial U$.  Three of the Kasner points
represent spacetimes which are isometric to a part of Minkowski
space, and so are called the flat Kasner spacetimes.  These are
labeled $T_i$, $i \in {1,2,3}$ (see figure~\ref{kasnercircle}).

\begin{table}
\caption{Strictly monotonic functions}

\vspace{15pt}
\setlength{\tabcolsep}{10pt}
\begin{tabular}{l|ccccc} \hline \hline
 & magnetic & magnetic & vacuum & vacuum & magnetic \\
 & Bianchi II$^+$ & Bianchi II$ ^-$ & Bianchi II$^+$  
 & Bianchi II$ ^-$ & Bianchi I \\ \hline
$L$ & decreasing & increasing & decreasing & increasing & constant \\
$W^+$ & increasing & indefinite & constant & decreasing & increasing \\
$W^-$ & indefinite &decreasing & increasing & constant & decreasing \\ 
\hline
\end{tabular}
\label{functions}
\vspace{9pt}
\end{table}

The orbit structure in the two-dimensional invariant sets of
the boundary, vacuum Bianchi~II and magnetic Bianchi~I, has been
known for some time~\cite[and references cited therein]{LKW}.
Each of these solutions is convergent in both time directions.
These are the transition solutions from one Kasner point to the
next in the BKL map.  There are two copies of vacuum Bianchi~II
in $\partial U$.  On one copy (which will be denoted vacuum
Bianchi~II$^+$) $N_+ = \sqrt{3} N_-$ and on the other (which will
be denoted vacuum Bianchi~II$^-$) $N_+ = -\sqrt{3} N_-$.  The
Kasner circle is the boundary of each of these three
two-dimensional open sets.  The union of these three sets and the
Kasner circle is the set $\cal A$ for magnetic Bianchi~VI$_0$.

The orbit structure can be seen by considering the behavior of
the following three functions on solutions.
\begin{eqnarray}
\label{monotonic}
L & = & { \Sigma_- \over 2 - \Sigma_+ } \nonumber \\
W^+ & = & { 1 + \Sigma_+ \over \sqrt{3} + \Sigma_-} \\
W^- & = & { 1 + \Sigma_+ \over -\sqrt{3} + \Sigma_-} \nonumber
\end{eqnarray}
Calculation of the time derivative of each of these functions,
using equations~\ref{equations} restricted to each set in turn,
shows that on each set one of these functions is constant on
solutions, one is strictly increasing on solutions and one is
strictly decreasing on solutions (see table \ref{functions}).
Thus the orbits are straight lines when projected into the
$\Sigma_+$--$\Sigma_-$ plane, and it can be determined which
Kasner point is in the $\alpha$-limit set and which Kasner
point is in the $\omega$-limit set of each solution.  The
$\omega$-limit set of a solution in magnetic Bianchi~I consists
of a point of ${\cal K}_1$, and the $\alpha$-limit set consists
of a point of ${\cal K}_2 \cup T_1 \cup {\cal K}_3$.  See
figure~\ref{kasnercircle} for the definition of ${\cal K}_i$. 
The $\omega$-limit set of a solution in vacuum Bianchi~II$^+$
consists of a point of ${\cal K}_2$ and the $\alpha$-limit set
consists of a point of ${\cal K}_3 \cup T_2 \cup {\cal K}_1$.
The $\omega$-limit set of a solution in vacuum Bianchi~II$^-$
consists of a point of ${\cal K}_3$ and the $\alpha$-limit set
consists of a point of ${\cal K}_1 \cup T_3 \cup {\cal K}_2$.

The polarized solutions in $U$, in which one of the degrees
of freedom of the gravitational field is absent and the
metric in a coordinate basis can be diagonalized \cite{diss},
form a two-dimensional invariant subset of $U$, defined by
$\Sigma_- = 0$ and $N_+ = 0$.  In \cite{LKW} it is shown that
the $\alpha$-limit set of each polarized solution in $U$ is
the heteroclinic cycle made up of the following orbits: \, the
Kasner points $T_1$ and $Q_1$ (see figure~\ref{kasnercircle}
for definition of $Q_i$), the single vacuum polarized
Bianchi~VI$_0$ orbit, $\Gamma_1$, whose $\alpha$-limit set
consists of the point $Q_1$ and whose $\omega$-limit set
consists of the point $T_1$, and the orbit in magnetic Bianchi~I,
$\Gamma_2$, whose $\alpha$-limit set is $T_1$ and whose
$\omega$-limit set is $Q_1$.  Thus the polarized solutions in $U$
are oscillatory toward the singularity.  They are not mixmaster.
Their evolution toward the singularity is not approximated by a
BKL sequence.  $\Gamma_2$ is a transition solution between two
sequential points in the BKL sequence, but $\Gamma_1$ is not. 
If a BKL sequence includes the point $T_i$ then it ends there.
Thus, a polarized solution in $U$ is an ``exceptional solution''
in this mixmaster class (see section 2).

In \cite{LKW} it is shown that the $\alpha$-limit set of a
non-polarized solution in $U$ must be contained in 
$\partial U$ by considering the function
\begin{equation}
\label{z}
Z = {3 \Sigma_-^2 + N_+^2 \over 3 N_-^2 - N_+^2}
\end{equation}
which is strictly decreasing on any non-polarized solution in $U$.

\section{Magnetic Bianchi VI$_0$ solutions are oscillatory}

That the magnetic Bianchi~VI$_0$ solutions are oscillatory
toward the singularity is shown by the following theorem.

\vspace{10pt} \noindent
{\it Theorem.} \, The $\alpha$-limit set of a spatially
homogeneous solution to Einstein's equation with Bianchi
type~VI$_0$ symmetry and a magnetic field orthogonal to
the two commuting Killing vector fields contains at
least two sequential Kasner points of the BKL map and the
orbit of the transition solution connecting them.

\vspace{10pt} \noindent {\it Proof of Theorem.} \,
The proof of the theorem will be given in the rest of this section.
The result has already been obtained for the polarized solutions
in $U$ \cite{LKW}, so it remains to obtain the result for the
non-polarized solutions in $U$.  It has also already been shown
that the $\alpha$-limit set of a non-polarized solution in $U$ is
contained in $\partial U$ \cite{LKW}.  Therefore the first part
of the proof of the theorem is an analysis of the orbit structure
in $\partial U$. It is shown that each limit
set of any solution in $\partial U$ contains a Kasner point.
From this it immediately follows that the $\alpha$-limit set of
a non-polarized solution in $U$ contains a Kasner point, because
a non-empty $\alpha$-limit set is a union of orbits along with
the limit sets of each orbit.  Therefore if an $\alpha$-limit
set is contained in $\partial U$ it must contain an entire orbit
in $\partial U$ and it must contain the orbit's limit sets.  Then,
through analysis of the orbit structure in the closure of $U$,
$\overline{U}$, in the neighborhood of a Kasner point, it is
shown that if the $\alpha$-limit set of a solution in $U$ contains
a Kasner point, it necessarily contains a second Kasner point.

\vspace{10pt} \noindent
{\it Proposition 1.} \, Each solution in the boundary of $U$ is
convergent to a Kasner point in both time directions.  If the
solution is not Kasner, then the Kasner point in its $\alpha$-limit
set is not the Kasner point in its $\omega$-limit set.

\vspace{10pt} \noindent {\it Proof of Proposition 1.} \, 
This result is already known for the Kasner points, the vacuum
Bianchi~II solutions and the magnetic Bianchi~I solutions that
lie in $\partial U$.

Consider the three-dimensional open invariant set, $I_3$,
made up of all the magnetic Bianchi~II$^+$
($N_+ = \sqrt{3} N_-$) orbits in $\partial U$.  The function
$L$ is strictly decreasing on solutions in $I_3$ and $W^+$ is
strictly increasing.  The existence of a function that is strictly
monotonic on solutions in an invariant set implies that the
limit sets of the solutions must be contained in the boundary
\cite[Proposition A1]{LKW}.  The boundary of $I_3$ is made up
of the union of all the Kasner points, the magnetic Bianchi~I
orbits and the vacuum Bianchi~II$^+$ orbits.  $L$ is strictly
decreasing on the latter solutions and $W^+$ is strictly
increasing on magnetic Bianchi~I solutions.  A function that
is strictly monotonic on a solution must be constant on each
limit set of that solution \cite{GD}, and a non-empty limit
set is the union of orbits.  Therefore each limit set of a
solution in $I_3$ must lie in the Kasner circle. For a given
value in the range of $L$ or $W^+$ there are at most two Kasner
points at which the function takes this value.  A limit set
must be connected.  Therefore each limit set of a solution
in $I_3$ must consist of a single Kasner point.  The Kasner
point in the $\alpha$-limit set of a solution is not the
the Kasner point in the $\omega$-limit set, since a function
that is strictly monotonic on a solution cannot have the
same value on both limit sets of the solution.

A similar argument, using the functions $L$ and $W^-$ (see
table \ref{functions}), shows that a solution in the set
$I_2$, made up of all the magnetic Bianchi~II$^-$
($N_+ = -\sqrt{3} N_-$) orbits in $\partial U$,
is convergent to a Kasner point in either time direction, and
that the Kasner point in the $\alpha$-limit set is not the
the Kasner point in the $\omega$-limit set.

Now consider the three-dimensional open invariant set, $I_1$,
made up of all the vacuum Bianchi~VI$_0$ orbits in
$\partial U$.  It was shown in \cite{GD} that each of
these solutions is convergent to a Kasner point in either
time direction.  In the present context, since the boundary
of $I_1$ is made up of the Kasner points and both sets of
vacuum Bianchi~II solutions, this result can be obtained
with the same reasoning used for $I_3$ by considering the
function $\Sigma_+$ which is strictly decreasing on
solutions in $I_1$ and is also strictly decreasing on both
sets of vacuum Bianchi~II solutions.  And again, the
existence of the monotonic function implies that the Kasner
point in the $\alpha$-limit set is not the Kasner point in
the $\omega$-limit set. \hfill $\Box$

It has now been shown that each limit set of any
solution in $\partial U$ contains a Kasner point, from which
it follows, as discussed above, that the $\alpha$-limit set
of a non-polarized solution in $U$ contains a Kasner point.
To show that this in turn implies that there is another
Kasner point in the $\alpha$-limit set, further analysis
of the orbit structure in $\overline U$ will be useful.

Figure \ref{kasnercircle} shows the sign of the eigenvalues
of the linearization of the differential equation about each
Kasner point \cite{LKW}.  At each Kasner point the eigenvector
corresponding to the eigenvalue $\lambda_H$ is tangent to
magnetic Bianchi~I while the eigenvectors corresponding to the
eigenvalues $\lambda_\pm$ are tangent to vacuum Bianchi~II$^\pm$.
The fourth eigenvalue, $\lambda_K$, vanishes at each Kasner point
and its eigenvector is tangent to the Kasner circle.  From the
number of eigenvalues which vanish, the number which are
positive and the number which are negative it follows that
each flat Kasner point has a three-dimensional center manifold
and a one-dimensional unstable manifold.  Each non-flat Kasner
point has a one-dimensional center manifold, a two-dimensional
unstable manifold and a one-dimensional stable manifold.  A
center manifold is not necessarily unique, while the stable and
unstable manifolds are unique.  If the orbit structure on a
center manifold is known, the orbit structure of the nonlinear
system near a Kasner point can be determined.  Let $O_C$ be the
intersection of a center manifold of the Kasner point with some
neighborhood, $N$, of the Kasner point in $\overline U$.
Let $j$, $k$ and $l$ be the dimensions of a center manifold,
the unstable manifold and the stable manifold, respectively.
Let $x=(x_1,...,x_j)$, $y=(y_1,...,y_k)$ and $z=(z_1,...,z_l)$.
Here $x_1,...x_j$ are coordinates on $O_C$, $y \in R^k$
and $z \in R^l$.  For some positive $\epsilon$, let
$$O = \{(x_1,...,x_j,y_1,...,y_k,z_1,...,z_l) : x \in O_C,
0 \leq y_i < \epsilon \mbox{\hspace{5pt}and\hspace{5pt}}
0 \leq z_i < \epsilon \}.$$
Let $x(\tau)$ satisfy the nonlinear system restricted to $0_C$.
Let $y(\tau) = y_0 \, e^\tau$ and $z(\tau) = z_0 \, e^{-\tau}$.
Then there exists an $\epsilon$ and an $N$ such that the
nonlinear system on $N$ is topologically equivalent to this
system on $O$ \cite[{\it Theorem 2.3}]{GD}.
(See also \cite[page 99]{AP}.)
Topological equivalence means that there is a homeomorphism,
$H:N \rightarrow O$, which maps orbits in $N$ onto orbits in
$O$ and preserves the time orientation on the orbits.

Let $q$ be a non-flat Kasner point.  Then the Kasner circle is
a center manifold of $q$.  Since the Kasner circle is made up
of equilibrium points, then in this case $x(\tau)$ as defined
above is independent of $\tau$.  The solutions on $O$ are
therefore known exactly.  Then the topological equivalence of
the nonlinear system on $N$ to this explicitly known system
on $O$ implies the following.  In $N$ the Kasner circle is
intersected by a foliation of three-dimensional leaves.  Any
orbit of the nonlinear system restricted to $N$ lies entirely
in one leaf.  (However, the intersection of an orbit in $U$ with
$N$ may have points on more than one leaf.)  One of the leaves
contains $q$, which is its own $\alpha$-limit set and its own
$\omega$-limit set.  In this leaf there is a one-dimensional stable
manifold of $q$, tangent to the eigenvector corresponding to the
negative eigenvalue.  The $\omega$-limit set of solutions on this
manifold is $q$.  There is a two-dimensional unstable manifold
of $q$, tangent to the two eigenvectors corresponding to the
positive eigenvalues.  The $\alpha$-limit set of solutions on
this manifold is $q$.  Now it will be shown that the stable and
unstable manifolds lie in $\partial U$ (and therefore not in $U$).

The one-dimensional stable manifold of $q$ has already been found.
If $q \in {\cal K}_1$ its stable manifold is an orbit of magnetic
Bianchi~I.  If  $q \in {\cal K}_2$ its stable manifold is an
orbit of vacuum Bianchi~II$^+$.  If $q \in {\cal K}_3$ its
stable manifold is an orbit of vacuum Bianchi~II$^-$.

Now consider the closure of $I_1$, $\overline I_1$.  In this
three-dimensional subsystem the eigenvalues of the linearization
of the differential equation about each Kasner point are
$\lambda_\pm$ and $\lambda_K$, with the same eigenvectors as in
the four-dimensional system.  If $q \in {\cal K}_1$ and $\tilde N$
is a neighborhood of $q$ in  ${\overline I}_1$, then the nonlinear
system restricted  to ${\tilde N} \cap N$ is topologically
equivalent to the system on $O$ with $l$ set to zero.  Therefore,
in ${\tilde N} \cap N$ the Kasner circle is intersected by
a foliation of two-dimensional  leaves, each of which is a
two-dimensional unstable manifold of the Kasner point that it
contains.  Since $\overline I_1$ is an invariant manifold, this
implies that the unstable manifold of $q$ lies in $\overline I_1$.

Similar reasoning shows that if $q \in {\cal K}_2$, its unstable
manifold lies in $\overline I_2$, while if $q \in {\cal K}_3$
its unstable manifold lies in $\overline I_3$.

Now let $S$ be a non-polarized solution in $U$ and let $\Lambda$
be its $\alpha$-limit set.  It was shown above that $\Lambda$
contains a Kasner point. Consider the case that $\Lambda$
contains $q$.  It has just been shown that $S$ is not in
either the stable or the unstable manifold of $q$.  Moreover,
analysis of the orbit structure in $O$ shows that no other
solutions there are convergent to $H(q)$ which shows that $S$
is not convergent to $q$.  But since $q$ is in $\Lambda$, then
there is a sequence of points on $S$ and in $N$, $S(\tau_n)$,
which converges to $q$, with $\tau_n < \tau_{n-1}$.  Let
$\Gamma_n$ be the orbit in $N$ on which $S(\tau_n)$ lies.
Note that $\Gamma_n$ need not be in the leaf containing $q$,
but the orbit structure in every leaf is the same and, in
addition, the sequence of $\Gamma_n$'s approaches the leaf
containing $q$.   Choose $\delta < \epsilon$ and define
$$O_\delta = \{(x,y_1,y_2,z) : x \in O_C,
0 \leq y_i \leq \delta \mbox{\hspace{5pt}and\hspace{5pt}}
0 \leq z \leq \delta \}.$$
Then each $H(\Gamma_n)$ has two points in the boundary of
$O_\delta$ in $H(N \cap U)$, one preceding $H(S(\tau_n))$
in time (because of the stable manifold) and one following
(because of the unstable manifold, but the preimage of this
point will have $\tau < \tau_{n-1})$).  This gives two sequences
of points,  one which converges to $H(p_s)$, with $p_s$ a
point on the stable manifold of $q$, and one which converges
to $H(p_u)$, with $p_u$ a point on the unstable manifold of
$q$.  Each point in each sequence is the image of a point
of the orbit of $S$.  Thus a point of the stable
manifold of $q$ and a point of the unstable manifold of
$q$ are in $\Lambda$.  For a more detailed argument, see
\cite[pages 33 and 34]{diss}.  The reasoning is taken from
\cite[Proof of {\it Theorem 4.3}]{GD} and the proof of a similar
result in \cite[Appendix]{RT}.  From this in turn it follows that
the entire stable manifold, its $\alpha$-limit set, an orbit of
the unstable manifold and its $\omega$-limit set must all lie
in $\Lambda$.  The $\alpha$-limit set of the stable manifold of $q$
is the Kasner point which follows $q$ in the BKL sequence.  The
stable manifold of $q$ is the transition solution between them.

The other possibility is that $\Lambda$ contains a flat Kasner
point $T_i$.  ${\overline I}_i$ is a center manifold of $T_i$
in $\overline U$.

\vspace{10pt} \noindent
{\it Proposition 2.} \, The $\omega$-limit set of any solution
in $I_i$ consists of the point $T_i$.

\vspace{10pt} \noindent {\it Proof of Proposition 2.} \,
It was already shown that the $\omega$-limit set of any solution
in $I_i$ consists of a single Kasner point.  The analysis
of the orbit structure in the neighborhood of a non-flat Kasner
point, $q$, showed that the stable manifold of $q$ is a single
orbit in one of the two-dimensional boundary sets, and that the
$\omega$-limit set of no other solution in $\overline U$
(besides $q$ itself) is the single point $q$.  The remaining
possibilities are the flat Kasner points, $T_i$.  

The function $W^+$ is strictly increasing on solutions in $I_3$.
Therefore the value of $W^+$ on the $\omega$-limit set of a
solution in $I_3$ must be greater than the infimum of $W^+$ on
$I_3$.  So $T_1$ cannot be in the $\omega$-limit set of a
solution in $I_3$.  The function $L$ is strictly decreasing on
solutions in $I_3$. Therefore the value of $L$ on the
$\omega$-limit set of a solution in $I_3$ must be less
than the supremum of $L$ on $I_3$.  So $T_2$ cannot be in
the $\omega$-limit set of a solution in $I_3$.  But the
$\omega$-limit set of a solution in $I_3$ consists of a single
Kasner point.  The only remaining possibility is $T_3$, so
$T_3$ must be the $\omega$-limit point of all solutions in $I_3$.

Similar reasoning shows that the $\omega$-limit point of all
solutions in $I_2$ is $T_2$.

To find out what is the case for $I_1$, first recall the
polarized case.  This one-dimensional set is a single orbit
whose $\alpha$-limit set is $Q_1$ and whose $\omega$-limit
set is $T_1$.  Now consider the (open and invariant) set
of non-polarized solutions in $I_1$. The function $Z$
(equation~\ref{z}) is bounded from below ($Z > 0$) and
it is strictly decreasing on solutions in this set.
Therefore it must approach a finite value on the
$\omega$-limit set of one of these solutions.  This is
impossible at $T_2$ and at $T_3$.  However, at $T_1$, if
$(3 \Sigma_-^2 + N_+^2) \rightarrow 0$ fast enough along
solutions compared to $3 N_-^2 - N_+^2$ then it is possible
for $Z$ to approach a finite value.  Since the $\omega$-limit
set must consist of a single Kasner point, and all the other
possibilities have been ruled out, then this must be the case.
\hfill $\Box$

\vspace{10pt} \noindent
{\it Proposition 3.} \, The $\alpha$-limit set of any solution
in $I_i$ consists of a point in ${\cal K}_i$.

\vspace{10pt} \noindent {\it Proof of Proposition 3.} \,
Given a point $q \in {\cal K}_i$, it has been shown that
the $\alpha$-limit sets of solutions on a two-dimensional
submanifold of $I_i$ consist of the point $q$.  It has been
ruled out that the $\alpha$-limit set of any other solution
in $\overline U$ can consist of the single point $q$.  It
has been shown that the $\alpha$-limit set of a solution
in $I_i$ consists of a single Kasner point.  The point $T_i$
is ruled out by the existence of a monotonic function on
solutions of $I_i$, since the $\omega$-limit set of solutions
of $I_i$ consists of the point $T_i$.  The remaining two flat
points, $T_j$, $j \neq i$ can be ruled out in each case by
considering $H_j(I_i \cap N_j)$, defined in the next paragraph.
\hfill $\Box$

Now consider the case that a flat Kasner point $T_i$ is in
$\Lambda$.  Let $O_C$ be the intersection of ${\overline I}_i$
with a neighborhood, $N_i \subset \overline U$, of $T_i$.
Let $x_1,x_2,x_3$ be coordinates on $O_C$.  Let
$$O_i = \{(x_1,x_2,x_3,y) : x \in O_C, 0 \leq y < \epsilon \}.$$
Let $x(\tau)$ satisfy the nonlinear system restricted to $0_C$
and $y(\tau) = y_0 \, e^\tau$.  Let $\epsilon$ and $N_i$ be such
that the nonlinear system on $N_i$ is topologically equivalent to
this system on $O_i$, with homeomorphism $H_i:N_i \rightarrow O_i$.
If $x$ is on the boundary of $I_i$ then $H_i^{-1}(x,y)$ is on the
boundary of $U$.  If $x \in I_i$ and $0 < y < \epsilon$ then
$H_i^{-1}(x,y) \in U$.  Since there exists a (small enough)
neighborhood of $T_i$ such that no solution in $I_i$ 
lies entirely within the neighborhood and since all solutions
in $I_i$ converge to $T_i$, $I_i$ plays the role
that the stable manifold played for the case that a non-flat
Kasner point is in $\Lambda$.  Then similar reasoning leads to
the result that the unstable manifold of $T_i$ and an orbit
other than $T_i$ in the closure of the center manifold along
with the limit sets of these two solutions must be in $\Lambda$.
$Q_i$ is the Kasner point preceding $T_i$ in the BKL sequence,
and the unstable manifold of $T_i$ is the transition solution
between them.  This ends the proof of the Theorem.

\section{From oscillatory to mixmaster}

It was found in the previous section that $\Lambda$ contains the
following: at least one non-flat Kasner point ($q_k$), the stable
manifold of $q_k$, its $\alpha$-limit set ($q_{k+1}$), an orbit
in the unstable manifold of $q_k$ and its $\omega$-limit set
(which is a single Kasner point).  If $q_{k+1}$ is a non-flat
point, then its stable manifold is also in $\Lambda$, and so on.
Let $q_k \rightarrow q_{k+1}$ stand for the union of $q_k$
and its stable manifold and $q_{k+1}$.  Then
\begin{equation}
\label{set}
q_k \rightarrow q_{k+1} \rightarrow q_{k+2}
\rightarrow q_{k+3} \rightarrow q_{k+4} \ldots
\end{equation}
is in $\Lambda$.  This BKL sequence of Kasner points can end at
a flat point, can be periodic, or can be non-ending and
non-periodic \cite{BKL2,HBC,cl}.

For this class of solutions to fit the standard picture of
mixmaster dynamics it must be the case that orbits in the
unstable manifold of $q_k$ that do not belong to $\cal A$ are
not in $\Lambda$, for generic solutions in $U$.  That is, there
should be no points of vacuum Bianchi~VI$_0$ and no points of
magnetic Bianchi~II in $\Lambda$.  A polarized solution in $U$
is exceptional.  There is an orbit, $\Gamma_1$ (see the end
of section 3), in its $\alpha$-limit set that does not belong
to $\cal A$. $\Gamma_1$ is the orbit of the vacuum polarized
Bianchi~VI$_0$ solution.  It will now be shown that no other
solutions in $U$ have this type of exceptional behavior.

\vspace{10pt} \noindent {\it Proposition 4.} \,
No points in the orbit of any vacuum Bianchi~VI$_0$ solution
are in the $\alpha$-limit set, $\Lambda$, of a non-polarized
magnetic Bianchi~VI$_0$ solution.

\vspace{10pt} \noindent {\it Proof of Proposition 4.} \,
The function $Z$ (equation~\ref{z}) is strictly decreasing on
non-polarized solutions in $U$, so $Z$ must be constant on
$\Lambda$.  $Z$ is also strictly decreasing on non-polarized
solutions in vacuum Bianchi~VI$_0$.  Since $\alpha$-limit sets are
the union of orbits then no point of a non-polarized vacuum
Bianchi~VI$_0$ orbit can be in $\Lambda$.  The polarized vacuum
Bianchi~VI$_0$ orbit is also ruled out since the value of
$Z$ at points of this orbit equals the infimum of $Z$ on
the set of non-polarized solutions in $U$.  \hfill $\Box$

This means that if some $q_k \in {\cal K}_1$ is in $\Lambda$ then
the only orbits in its unstable manifold that can be in $\Lambda$
are the two that are in vacuum Bianchi~II. These are precisely
the orbits of the transition solutions between $q_{k-1}$ and $q_k$.
(Given a non-flat Kasner point $q$, there are two Kasner points
which can immediately precede $q$ in the BKL sequence.)

Two possibilities come to mind for showing that no points of
magnetic Bianchi~II are in $\Lambda$, for generic solutions
in $U$.  One possibility is more careful analysis of the orbit
structure in $\overline U$ in the neighborhood of a Kasner point.
Another is the following.  If two functions analogous to $Z$
could be found, both of which were strictly monotonic on (perhaps
generic) solutions in $U$, one also monotonic on (perhaps generic)
solutions in magnetic Bianchi~II$^+$ and the other monotonic on
(perhaps generic) solutions in magnetic Bianchi~II$^-$, then the
same result just obtained for $q_k \in {\cal K}_1$ would be
obtained for $q_k \in {\cal K}_2$ and for $q_k \in {\cal K}_3$.
Only two orbits in the unstable manifold of $q_k$ would be left
as possibilities for being in $\Lambda$.  These are the orbits
of the transition solutions of the BKL map between the two
possible $q_{k-1}$'s and $q_k$.  One or the other of these
orbits would have to be in $\Lambda$.  Therefore a sequence
of the following form would have to be in the $\alpha$-limit
set of generic magnetic Bianchi~VI$_0$ solutions.
\begin{equation}
\label{sequence}
\ldots q_{k-2} \rightarrow q_{k-1} \rightarrow q_{k}
\rightarrow q_{k+1} \rightarrow q_{k+2} \ldots
\end{equation}
This sequence is either periodic or infinite, since, while the
domain of the BKL map is the set of all Kasner points except for
the three flat Kasner points, the range is the set of all Kasner
points.  Therefore the sequence can end (a point is in the range
that is not in the domain), but it cannot begin (every point that
is in the domain is in the range).

Showing that no points of magnetic Bianchi~II belong to
$\Lambda$ would, furthermore, imply that if the
sequence~\ref{sequence} ends at a $T_i$, that is, if $T_i$ is in
$\Lambda$, then so is a non-flat Kasner point in any neighborhood
of $T_i$.  As of now, this result can only be obtained for $T_1$.

\vspace{10pt} \noindent {\it Proposition 5.} \,
If $T_1$ is in the $\alpha$-limit set, $\Lambda$, of
a non-polarized magnetic Bianchi~VI$_0$ solution, then
for any neighborhood, $N_\epsilon$, of $T_1$, no matter how
small, there is a non-flat Kasner point $q$ such that
$q \in N_\epsilon \cap \Lambda$.

\vspace{10pt} \noindent {\it Proof of Proposition 5.} \,
In the proof of the theorem, it was found that if
$T_1 \in \Lambda$ then an orbit in ${\overline I}_1$ is also
in $\Lambda$. But it has now been shown that no points of vacuum
Bianchi~VI$_0$ are in $\Lambda$.  Therefore, an orbit in vacuum
Bianchi~II is in $\Lambda$.  Furthermore, the reasoning in the
proof of the theorem shows that for any neighborhood $N_\delta$
of $T_1$ that is small enough, there is a vacuum Bianchi~II
orbit in $\Lambda$ which intersects the boundary of $N_\delta$
in $\overline U$.  But then the limit sets of the orbit must
also be in $\Lambda$.  So for any neighborhood, $N_\epsilon$,
of $T_1$, choose $N_\delta$ with the following properties:

\vspace{10pt} \noindent i. $N_\delta$ is a neighborhood of $T_1$
in $\overline U$.

\noindent ii. No solutions in $I_1$ are entirely contained
in $N_\delta$ and the unstable manifold of $T_1$ is not
entirely contained in $N_\delta$.

\noindent iii. $N_\delta \subset N_1$, where $N_1$ is a
neighborhood of $T_1$ which is topologically equivalent
to $O_1$, as given at the end of section 4.

\noindent iv. The limit points of all vacuum
Bianchi~II orbits which intersect ${\overline N}_\delta$
are contained in $N_\epsilon$.

\vspace{10pt} \noindent
None of these limit points is $T_1$.
All of these limit points are non-flat Kasner points.   
At least one of these limit points is in $\Lambda$.
\hfill $\Box$

\section{Strong cosmic censorship}

While the results of this paper do not give a complete
characterization of the $\alpha$-limit set of each magnetic
Bianchi~VI$_0$ solution, they are sufficient for showing that
none of these solutions has an extension past the initial
singularity.  That this is the case follows from the
unboundedness of the Kretschmann scalar,
$R_{\alpha \beta \gamma \delta} R^{\alpha \beta \gamma \delta}$,
in a neighborhood of the singularity.  That the Kretschmann
scalar is unbounded in a neighborhood of the singularity follows
from the unboundedness of the expansion scalar in a neighborhood
of the singularity and the presence of a non-flat Kasner point
in the $\alpha$-limit set (so there is a sequence of points on
the orbit of any magnetic Bianchi~VI$_0$ which both ``goes to
the singularity'' and converges to a non-flat Kasner point).
The argument is the same as in the proof of {\it Theorem 5.1}
in \cite{GD}.  See also \cite[equations (3.5) and (3.6)]{GD}.  The
ratio between the Kretschmann scalar and the fourth power of
the expansion scalar can be expressed as a polynomial in
$\Sigma_+$, $\Sigma_-$, $N_+$, $N_-$ and $H$.  This ratio is
nonvanishing at a non-flat Kasner point, and the result follows.

The same argument shows that the Kretschmann scalar is unbounded
in a neighborhood of the singularity in each of the solutions
considered in this paper which has a non-flat Kasner point in
its $\alpha$-limit set, and so such a solution cannot be
extended past the singularity.  This is the case for each
vacuum Bianchi~VI$_0$ solution and each of the magnetic
Bianchi~II solutions considered in this paper.  It is also
the case for ``most'' of the vacuum Bianchi~II solutions,   
the magnetic Bianchi~I solutions and the Kasners themselves.

The exceptions are the flat Kasner spacetimes, the rotationally
symmetric vacuum  Bianchi~II spacetimes and the rotationally
symmetric magnetic Bianchi~I spacetimes.  The $\alpha$-limit set
of each of these consists of a single flat Kasner point and the
argument just given is inconclusive in these cases (the ratio
between the Kretschmann scalar and the fourth power of the
expansion scalar vanishes).  It turns out that each of these
spacetimes can, in fact, be extended past the initial singularity.
It is already known that the flat Kasner spacetimes and the
rotationally symmetric vacuum Bianchi~II spacetimes \cite{siklos}
can be extended.  An extension for the rotationally symmetric
magnetic Bianchi~I spacetimes is given in the appendix.

\section{Conclusion}
It has been shown in this paper that any spatially homogeneous
solution to Einstein's equation of Bianchi type~VI$_0$ with source
consisting of a magnetic field orthogonal to the two commuting
Killing vector fields has an oscillatory approach to the initial
singularity and has no extension past the singularity.  In the
formulation which has been used, other spatially homogeneous
solutions lie on the boundary of magnetic Bianchi~VI$_0$.  All
of the boundary solutions are convergent rather than oscillatory,
and most are inextendible past the initial singularity.  The only
ones which are extendible are the rotationally symmetric vacuum
Bianchi~II solutions, the rotationally symmetric magnetic Bianchi~I
solutions and the flat Kasners.  Since these are non-generic in the
class, strong cosmic censorship is supported.

The methods used here to characterize the approach to the singularity
are those of dynamical systems analysis.  One would like to obtain
a rigorous proof that generic non-polarized magnetic Bianchi~VI$_0$
solutions are mixmaster.  This result could be obtained by finding
two functions which are strictly monotonic on all or at least generic
magnetic Bianchi~VI$_0$ solutions as formulated here and also, in
turn, on the two classes of magnetic Bianchi type~II solutions which
lie on the boundary.  The existence of such functions would exclude
these solutions of the boundary from the $\alpha$-limit sets of
magnetic Bianchi~VI$_0$ solutions.  Alternatively, one could try
to show that this class is mixmaster by other methods.  It appears
that such a result has been recently obtained by Ringstr\"{o}m
for solutions of vacuum Bianchi types~VIII and IX \cite{ring}.

It is hoped that this work will contribute to the understanding
of the nature of cosmological singularities in classical
General Relativity.  In particular, the validity in the generic
case of the description of the approach to the cosmological
singularity given by Belinskii, Khalatnikov and Lifshitz
is still an open question.  Whether or not the methods
of dynamical systems analysis turn out to be directly useful
in studying this question in spatially inhomogeneous classes
of cosmological solutions to Einstein's equation, they have
added to our understanding of the spatially homogeneous
case, which gives a surer foundation for further study.

\section*{Acknowledgements}

I would like to thank Alan Rendall for suggesting this problem
and Jim Isenberg for interesting and helpful discussions
and for comments on a draft of this paper.
This work was supported by a U. S. Federal Department of
Education Fellowship through the University of Oregon.

\section*{Appendix}

The metric of a rotationally symmetric Bianchi~I spacetime
with a magnetic field orthogonal to the plane of rotational
symmetry can be written as follows \cite[Appendix C, Case 6]{LKW}.
\begin{equation}
g=-A^2 \, dt^2 + t^2 A^{-2} \, dx^2 + A^2 \, dy^2 + A^2 \, dz^2
\end{equation}
with $A = 1 + {1 \over 4} B^2 t^2$.  The Maxwell tensor is
$F = B \, (dy \otimes dz - dz \otimes dy)$, with $B$ a constant.
In the direction of the singularity ($t \rightarrow 0$) the
generalized Kasner exponents converge to $(1,0,0)$.  This is
the point $T_1$ in the formulation used in the present paper.
In the opposite time direction ($t \rightarrow \infty$) the
generalized Kasner exponents converge to
$(-{1 \over 3}, {2 \over 3}, {2 \over 3})$.  This is
the point $Q_1$ in the formulation used in the present paper.
These solutions were given by Rosen \cite{rosen}
and studied by Jacobs \cite{jacobs}, but the author
is not aware of a published extension of these spacetimes.

Here is a coordinate transformation that allows each of
these spacetimes (inequivalent spacetimes result from
different values of the constant $|B|$) to be extended.
\begin{equation}
\tilde{t} = t \, \cosh x \hspace*{40pt} \tilde{x} = t \sinh x
\end{equation}
In the original spacetime, $t^2 = \tilde{t}^2 - \tilde{x}^2$.
But if a function $D$ is defined by
$D = \tilde{t}^2 - \tilde{x}^2$, and if the function $A$ is now
defined $A = 1 + {1 \over 4} B^2 D$, then the transformed metric
\begin{equation}
g=A^2 \, (-d\tilde{t}^2 + \, d\tilde{x}^2)
+(-B^2 + B^4 D {32 - B^4 D^2 \over 256 A^2}) (\tilde{x} \, d\tilde{t}
- \tilde{t} \, d\tilde{x})^2 + A^2 \, dy^2 + A^2 \, dz^2
\end{equation}
is a solution to Einstein's equation in the region $A > 0$.  This
is an extension of the original solution.  The region defined by
$0 \geq D > -{4 \over B^2}$ is not in the original spacetime.  The
singularity at $A=0$ is a curvature singularity.  Both the Kretschmann
scalar and $R_{\alpha \beta} R^{\alpha \beta}$ are unbounded as
$A \rightarrow 0$.
\begin{equation}
R_{\alpha \beta \gamma \delta} R^{\alpha \beta \gamma \delta}
=(20 - 6 B^2 D + {3 \over 4} B^4 D^2) {B^4 \over A^8} \hspace*{20pt}
\: \: R_{\alpha \beta} R^{\alpha \beta}= {4 B^4 \over A^8}
\end{equation}
$\partial_{\tilde{t}}$ is everywhere timelike, but
$\partial_{\tilde{x}}$ is not everywhere spacelike.
However, $g_{\tilde{t} \tilde{t}} \, g_{\tilde{x} \tilde{x}}
- g_{\tilde{t} \tilde{x}}^2 = -1$.  Thus the two-dimensional
surface defined by constant $y$ and $z$ is everywhere timelike.

The singularity is timelike.

Along the $\tilde{x}$-axis the spacetime becomes singular a finite
distance from the origin in either direction.

\end{document}